\documentclass[aps,prb,twocolumn,superscriptaddress,floatfix,showpacs,10pt,letterpaper]{revtex4}
\usepackage{graphicx}% Include figure files
\usepackage{dcolumn}% Align table columns on decimal point
\usepackage{bm}% bold math
\usepackage{subfigure}
\usepackage{amssymb,amsmath}
\usepackage{relsize}
\usepackage{lmodern}
\usepackage{slantsc}
\usepackage{scalefnt}
\usepackage{tikz}
\usetikzlibrary{arrows,calc,scopes}
\usetikzlibrary{shapes.geometric}
\allowdisplaybreaks[1]

\usepackage[colorlinks,citecolor=blue]{hyperref}
\newcommand{\be}{\begin{equation}} \newcommand{\ee}{\end{equation}}
\newcommand{\bse}{\begin{subequations}}\newcommand{\ese}{\end{subequations}}
\newcommand{\bpm}{\begin{pmatrix}} \newcommand{\epm}{\end{pmatrix}}
\newcommand{\bmm}{\begin{matrix}} \newcommand{\emm}{\end{matrix}}

\newcommand{\Z}{\mathbb{Z}} 
\newcommand{\R}{\mathbb{R}}

 \renewcommand{\t}[1]{\tilde{#1}}

\newcommand{\e}{\hspace{1pt}\mathrm{e}}

\newcommand{\Rf}[1]{Ref.~\onlinecite{#1}}

 \newcommand{\eqn}[1]{eqn.~(\ref{#1})}

\newcommand{\<}{\langle} \renewcommand{\>}{\rangle}

\newcommand{\ie}{{\it ie~}}  \newcommand{\etc}{{\it
etc~}}

\newcommand{\al}{\alpha} \newcommand{\bt}{\beta} \newcommand{\del}{\delta}

 \newcommand{\cH}{ {\cal H} }

\usepackage{varioref}
\begin{document}

%\preprint{APS/123-QED}

\begin{titlepage}

\title{Universal symmetry-protected topological invariants\\
for symmetry-protected topological states}
% Force line breaks with \\

\author{Ling-Yan Hung}\email{lhung@physics.harvard.edu}
\affiliation{Department of Physics, Harvard University, Cambridge MA 02138}
\author{Xiao-Gang Wen}
\affiliation{Perimeter Institute for Theoretical Physics, 31 Caroline St N,
Waterloo, ON N2L 2Y5, Canada}
\affiliation{Department of Physics, Massachusetts Institute of Technology,
Cambridge, Massachusetts 02139, USA}
\affiliation{Collaborative Innovation Center of Quantum Matter, Beijing, China}

\date{\today}% It is always \today, today,

\begin{abstract}
Symmetry-protected topological (SPT) states are short-range entangled states
with a symmetry $G$.  They belong to a new class of quantum states of matter
which are classified by the group cohomology $\cH^{d+1}(G,\R/\Z)$ in
$d$-dimensional space.  In this paper, we propose a class of symmetry-protected
topological invariants that may allow us to fully characterize SPT states with
a symmetry group $G$ (\ie allow us to measure the cocycles in
$\cH^{d+1}(G,\R/\Z)$ that characterize the SPT states). We give an explicit and
detailed construction of symmetry-protected topological invariants for 2+1D SPT
states. Such a construction can be directly generalized to other dimensions.

\end{abstract}

\maketitle
\end{titlepage}

{\small \setcounter{tocdepth}{1} \tableofcontents }

\section{Introduction}

Quantum states of matter have shown a lot of fascinating properties which
require a completely new way of understanding.  Recent study of long range
quantum entanglement\cite{CGW1038} (as defined through local unitary (LU)
transformations.\cite{LW0510,VCL0501,V0705}) reveal a direct connection between
entanglement and gapped phases of quantum matter.  The notion of long range
entanglement leads to a more general and more systematic picture of gapped
quantum phases and their phase transitions.\cite{CGW1038} For gapped quantum
systems without any symmetry, their quantum phases can be divided into two
classes: short range entangled (SRE) states and long range entangled (LRE)
states.

SRE states are states that can be transformed into direct product states via LU
transformations. 
LRE states are states that cannot be transformed into  direct product states
via LU transformations.  There are many types of LRE states that cannot be
transformed into each other via the LU transformations.  Those different types
of LRE states are nothing but the topologically ordered phases.  Fractional
quantum Hall states\cite{TSG8259,L8395}, chiral spin
liquids,\cite{KL8795,WWZ8913} $Z_2$ spin liquids,\cite{RS9173,W9164,MS0181}
non-Abelian fractional quantum Hall states,\cite{MR9162,W9102} \etc are
examples of topologically ordered phases.  

For gapped quantum systems with symmetry, the structure of phase diagram is
even richer.  SRE states now can belong to different phases. The Landau
symmetry breaking states belong to this class of phases, where different states
are characterized by their different symmetries.  However, even if there is no
symmetry breaking, the SRE states that have the same symmetry can still belong
to different phases.\cite{CLW1141,CGL1172,CGL1204}  The 1D Haldane phases for
spin-1 chain\cite{H8364,AKL8877} and topological
insulators\cite{KM0501,BZ0602,KM0502,MB0706,FKM0703,QHZ0824} are examples of
non-trivial SRE phases that do not break any symmetry.  Those phases are beyond
Landau symmetry breaking theory since they do not break any symmetry.  Those
phases are called Symmetry Protected Topological (SPT) phases, which are under
intense study
recently.\cite{LS0903,LV1219,LW1224,CW1217,HW1267,HW1232,VS1258,X1299,LL1209,HW1227,LL1263,YW1221,OCX1226,XS1372,WS1334,BCF1372,CG1303,CWL1321,CLV1301,YW1372,MKF1335,GL1369,LV1334} 

We know that topological order\cite{Wtop,WNtop,Wrig} (\ie patterns of
long-range entanglement) cannot be characterized by the local order parameters
associated with the symmetry breaking.  We have to use topological probes to
characterize/define topological order. It appears that we only need two
topological probes to characterize/define 2+1D topological orders:
(a) the robust ground state degeneracy that depend on the spatial
topologies\cite{Wtop,WNtop} but cannot be lifted by any small perturbations,
(b) the quantized non-Abelian geometric phases from deforming the degenerate
ground states.\cite{Wrig,KW9327} Using  ground state degeneracy and non-Abelian
geometric phases to  characterize/define topological order is just like using
zero-viscosity and quantized vorticity to characterize/define superfluid order.
In some sense, the robust ground state degeneracy and the non-Abelian geometric
phases (that generate modular representation of the degenerate ground states)
can be viewed as a type of ``topological order parameters'' for topologically
ordered states.  Those ``topological order parameters'' are also referred to as
topological invariants of topological order.

With the above understanding of topological order, we like to ask: What are the
``topological order parameters'' or the symmetry-protected topological
invariants that can be used to characterize/define SPT states?  One way to
characterize SPT states is to create a boundary, and then study the boundary
properties.\cite{CLW1141,LG1220,LW1224,CW1217,VS1258,WS1334,BCF1372,GL1369}
This approach is very practical since the boundary can be probed in
experiments.  But it is not convenient theoretically, since the different ways
to create the boundary can lead to different  boundary properties, even for the
same bulk SPT state.  Another way to characterize SPT states is to gauge the
on-site symmetry\cite{LG1220} and use the introduced gauge field as an
effective probe for the SPT order.\cite{W1375}. (See also \Rf{Santos:2013uda}.) 
This will be the main theme of
this paper. In \Rf{W1375}, many SPT invariants are discussed and constructed
based on the structure of the group cohomology class that described the SPT
states. However, the construction in \Rf{W1375} is not systematic and we often
fail to find SPT invariants that fully characterize the SPT state.  In this
paper, we will try to systematically construct  SPT invariants that can fully
characterize the SPT state.

We find that we can use the introduced gauge field in a SPT state to
``simulate'' the degenerate ground states of intrinsic topological order.  We
can even  use the  introduced gauge field in a SPT state to ``simulate'' the
quantized non-Abelian geometric phases describe by a unitary matrix $U$ from
deforming the ``simulated'' degenerate ground states.  
We propose an easy way to compute such non-Abelian geometric phases:
the matrix elements of $U$ can be computed from the overlap
of a ``simulated'' degenerate ground state and its twist.
For example, in 2+1D, we have
\begin{align}
\<\al| \hat U |\bt\>
= \text{e}^{-L^2/\xi^2+o(1/L)}
 U_{\al\bt} 
\end{align}
where $L^2$ is the area of the system, $|\al\>,|\bt\>$ are the ``simulated''
degenerate ground states, and $\hat U$ is the operator that generate the twist
(see \eqn{tS}, \eqn{tT}, and \eqn{ST} for more details).  The coefficient
$1/\xi^2$ is not universal, while the factor $U_{\al\bt}$,
we believe, is universal.\cite{MW13,HW13} We have a similar result for higher
dimensions. 

We find a direct relation between the quantized non-Abelian geometric phases
and the topological partition function on space-time which can be an arbitrary
fiber bundle over $S^1$.  The different choices of the twist $\hat U$
correspond to different fiber bundle.  This makes us to believe\cite{KW13} that
the quantized non-Abelian geometric phases from deforming the simulated
degenerate ground states are the ``topological order parameters'' or the SPT
invariants that can be used to fully characterize/define SPT states.

\section{Universal topological invariants of SPT orders}

\subsection{Symmetry twist}

\begin{figure}[tb] 
\begin{center} 
\includegraphics[scale=0.4]{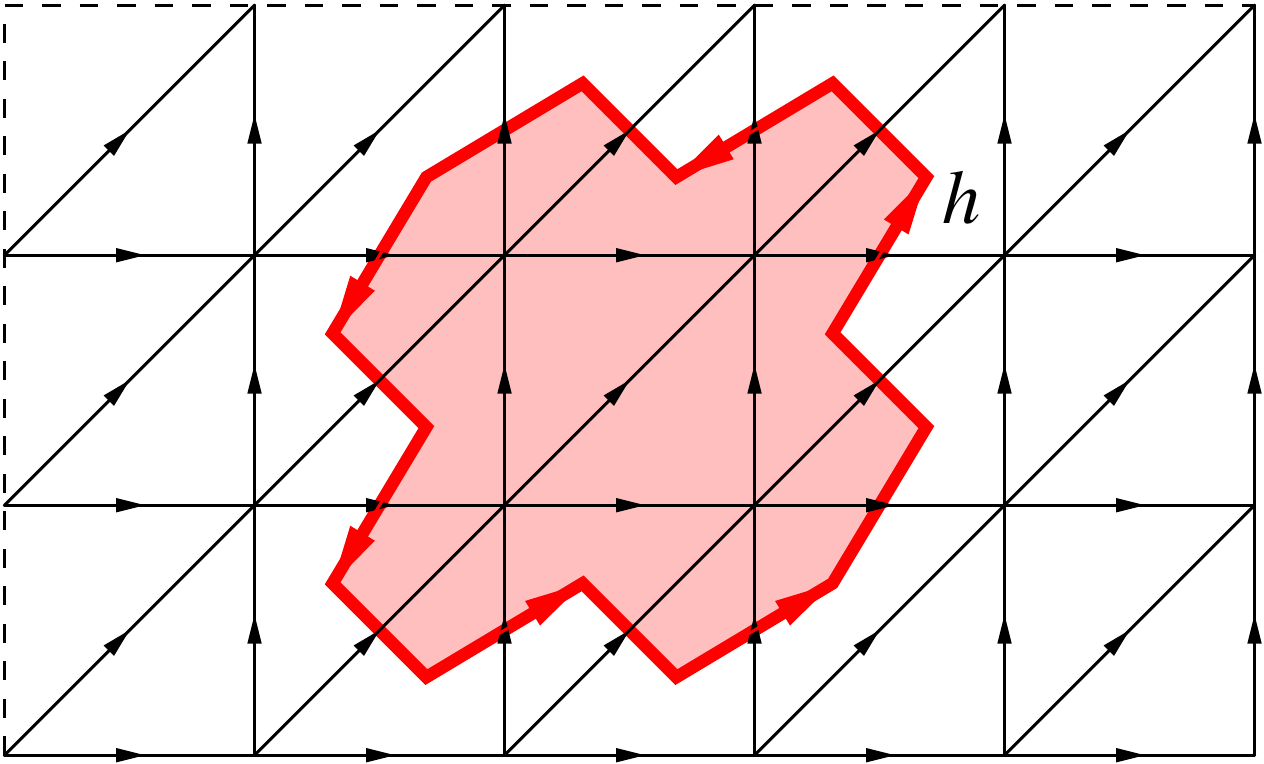} \end{center}
%Fig. 1
\caption{ (Color online) 
A 2D lattice on a torus.  A $h\in G$ transformation is performed on the sites in
the shaded region.  The $h$ transformation changes the Hamiltonian term on
the triangles $(ijk)$ across the boundary (the loop) from $H_{ijk}$ to $H'_{h,ijk}$.
} 
\label{ltrans} 
\end{figure}

In order to use  introduced gauge field to  ``simulate''
the degenerate ground states, let us introduce the notion
of ``symmetry twist''.
We first assume that the 2D lattice Hamiltonian for a SPT state 
with symmetry $G$ has a
form (see Fig. \ref{ltrans})
\begin{align}
 H=\sum_{(ijk)} H_{ijk},
\end{align}
where $\sum_{(ijk)}$ sums over all the triangles in Fig. \ref{ltrans} and
$H_{ijk}$ acts on the states on site-$i$, site-$j$, and site-$k$:
$|g_ig_jg_k\>$.  (Note that the states on site-$i$ are labeled by $g_i \in
G$.)  $H$ and $H_{ijk}$ are invariant under the global $G$ transformations.

Then we perform a  $G$ transformation, $h \in G$, only in the shaded region in
Fig.  \ref{ltrans}. Such a transformation will change $H$ to $H'$.  However,
only the Hamiltonian terms on the triangles $(ijk)$ across the boundary are
changed from $H_{ijk}$ to $H'_{h,ijk}$.  Since the  $g\in G$ transformation is an
unitary transformation, $H$ and $H'$ have the same energy spectrum.  In other
words the boundary in Fig.  \ref{ltrans} (described by $H'_{h,ijk}$'s) do not
cost any energy.

Now let us consider a Hamiltonian on a lattice with a ``loop'' (see Fig.
\ref{ltrans})
\begin{align} \label{loopH}
H^\text{gauged}_h= { \sum_{(ijk)}}' H_{ijk} +{\sum_{(ijk)}}^\text{loop} H'_{h,ijk}
\end{align}
where $\sum'_{(ijk)}$ sums over the triangles not on the loop and
$\sum^\text{cut}_{(ijk)}$ sums over the triangles that are divided into
disconnected pieces by the loop.  
We note that the loop carries no energy. 
The Hamiltonian $H^\text{gauged}_h$ defines the symmetry twist generated by $h\in G$.
We like to point out that the above procedure to obtain $H^\text{gauged}_h$ is actually
the ``gauging'' of the $G$ symmetry.  $H^\text{gauged}_h$ is a gauged Hamiltonian that
contains a locally flat gauge configuration. 

\subsection{Simulating nearly degenerate ground states}

\begin{figure}[tb] 
\begin{center} 
\includegraphics[scale=0.4]{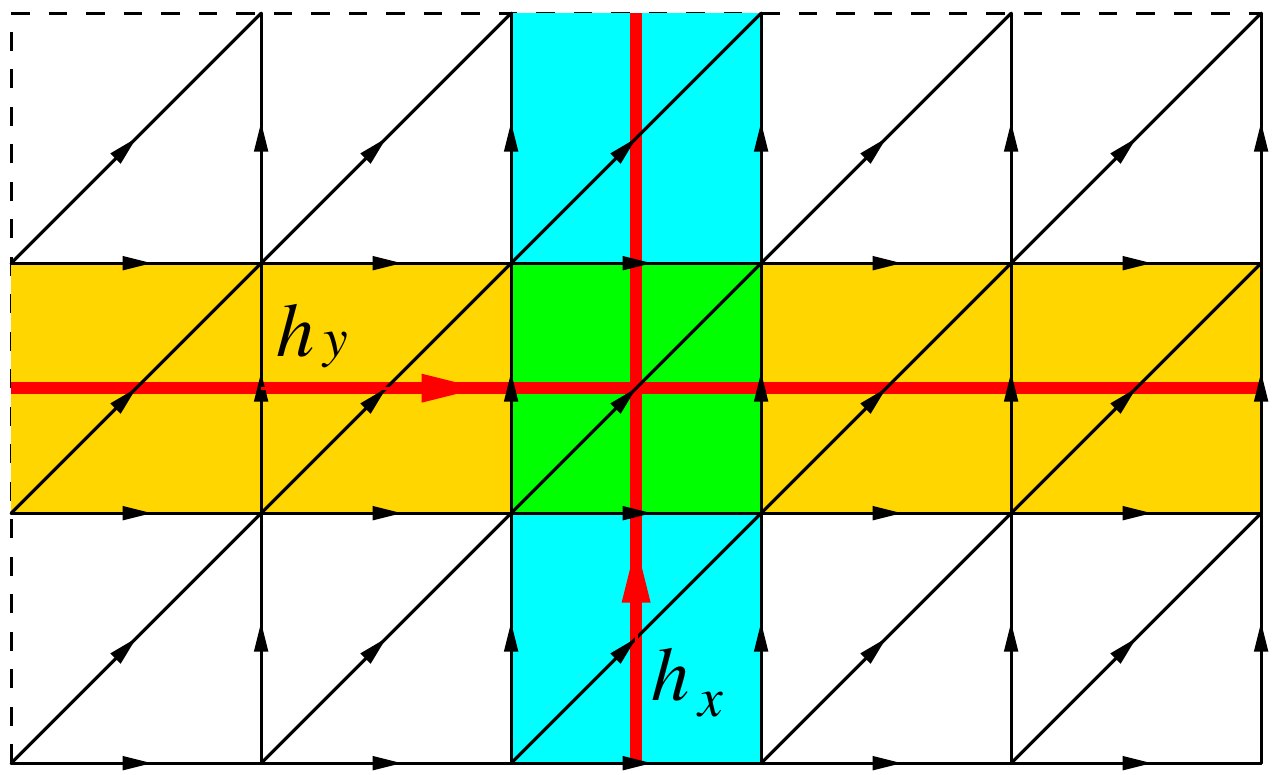} \end{center}
%Fig. 2
\caption{ (Color online) 
The Hamiltonian $H^\text{gauged}_{h_x,h_y}$ with two symmetry twists $h_x$ and
$h_y$ along the loops in $y$- and $x$- directions respectively.  The shaded
triangles $(ijk)$ across the the loop contain Hamiltonian terms $H'_{h_x,ijk}$
or $H'_{h_y,ijk}$.
} 
\label{hxhy} 
\end{figure}

The 2+1D topologically ordered states usually have topologically robust
nearly degenerate ground states on torus.  Such  nearly degenerate ground states can be
simulated in SPT states via the symmetry twists discussed above.

To simulate the nearly degenerate ground states on torus, we consider a SPT state on
torus with two symmetry twists $h_x$ and $h_y$ along the loops in $y$- and $x$-
directions respectively (see Fig. \ref{hxhy}).  The resulting Hamiltonian is
denoted as $H^\text{gauged}_{h_x,h_y}$, and its unique ground state
is denoted as $|\Psi_{h_x,h_y}\>$.

We note that the loops in $y$- and $x$- directions intersect.
In order that the  intersecting point does not cost any energy
(\ie in order for the gauged Hamiltonian to describe a locally flat
gauge configuration), we require that
\begin{align}
 [h_x,h_y]=0.
\end{align}
$|\Psi_{h_x,h_y}\>$ with commuting $h_x,h_y$ simulates the nearly degenerate
group states.

A symmetry twist locally breaks the global symmetry, and group action
can be defined on them. In our construction, a twist characterized by 
a group element $h$ is  transformed to $h\to g h g^{-1}$. i.e. The group
acts by conjugation. Therefore, twists connected by orbits of the group action
falls into the same conjugacy class. This is analogous to the usual quotient group
$G/H$ classification of defects, where $H$ is a normal subgroup keeping the defect invariant.
It is then clear that the $G$ symmetry implies that the ground states 
$|\Psi_{h_x,h_y}\>$ and $|\Psi_{gh_xg^{-1},gh_yg^{-1}}\>$ of
$H^\text{gauged}_{h_x,h_y}$ and $H^\text{gauged}_{gh_xg^{-1},gh_yg^{-1}}$ respectively, 
connected by a group action $I(g), g\in G$, i.e.
\be
I(g)|\Psi_{h_x,h_y}\>  = I_{(gh_xg^{-1},gh_yg^{-1}), (h_x,h_y)}|\Psi_{gh_xg^{-1},gh_yg^{-1}}\>,
\ee 
for some characteristic $U(1)$ phase $I_{(gh_xg^{-1},gh_yg^{-1}), (h_x,h_y)}$
have exactly the same energy.  
In an intrinsic topological order constructed from the gauged theory with the
same group $G$, it is the  equivalence class
$\{|\Psi_{gh_xg^{-1},gh_yg^{-1}}\>| g\in G \}$ that corresponds to a single
nearly degenerate group state.

If $G$ is Abelian and finite, each equivalent class contains only one state
$|\Psi_{h_x,h_y}\>$, and the total number of equivalent classes is $|G|^2$,
where $|G|$ is the number of elements in $G$.  This agrees with the topological
ground state degeneracy of $G$-gauge theory on torus which is also $|G|^2$ if
$G$ is Abelian.

\subsection{Simulate the non-Abelian geometric phases}

\begin{figure}[tb] 
\begin{center} 
\includegraphics[scale=0.4]{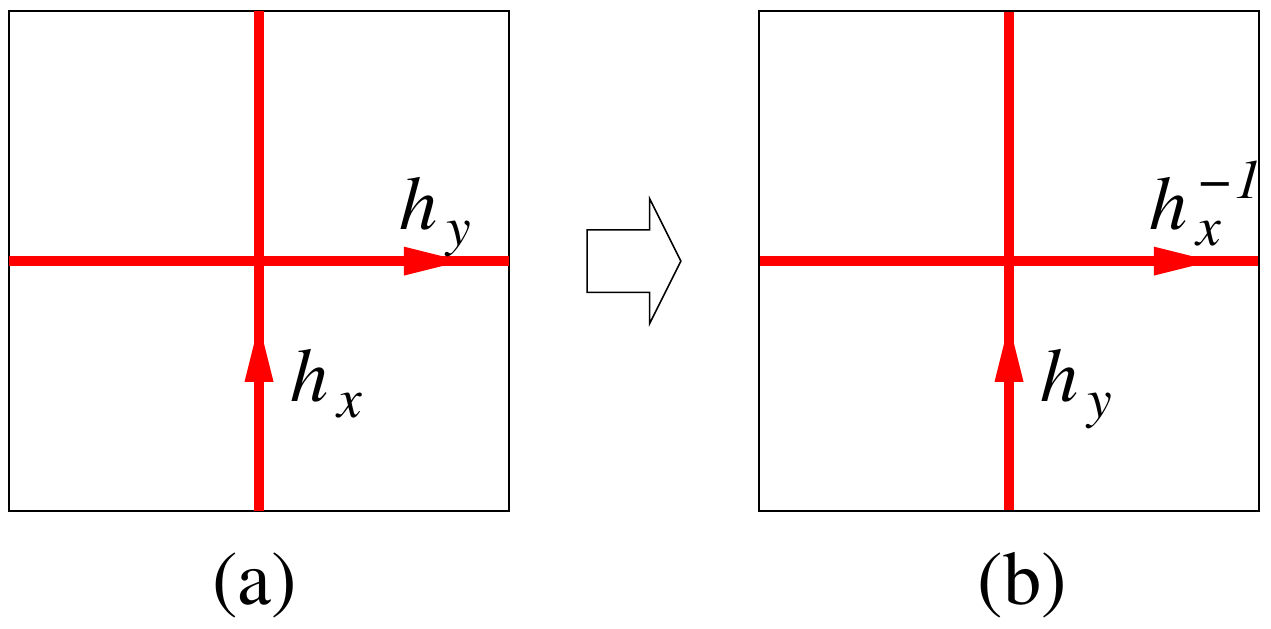} \end{center}
%Fig. 3
\caption{ (Color online) 
(a) A system on a torus with two symmetry twists in $x$- and $y$-directions.
Note that the  torus has the same size $L$ in $x$- and $y$-directions.
(b) The $S$-transformation of the torus, and the
resulting new symmetry twists.
} 
\label{Smov} 
\end{figure}

\begin{figure}[tb] 
\begin{center} 
\includegraphics[scale=0.4]{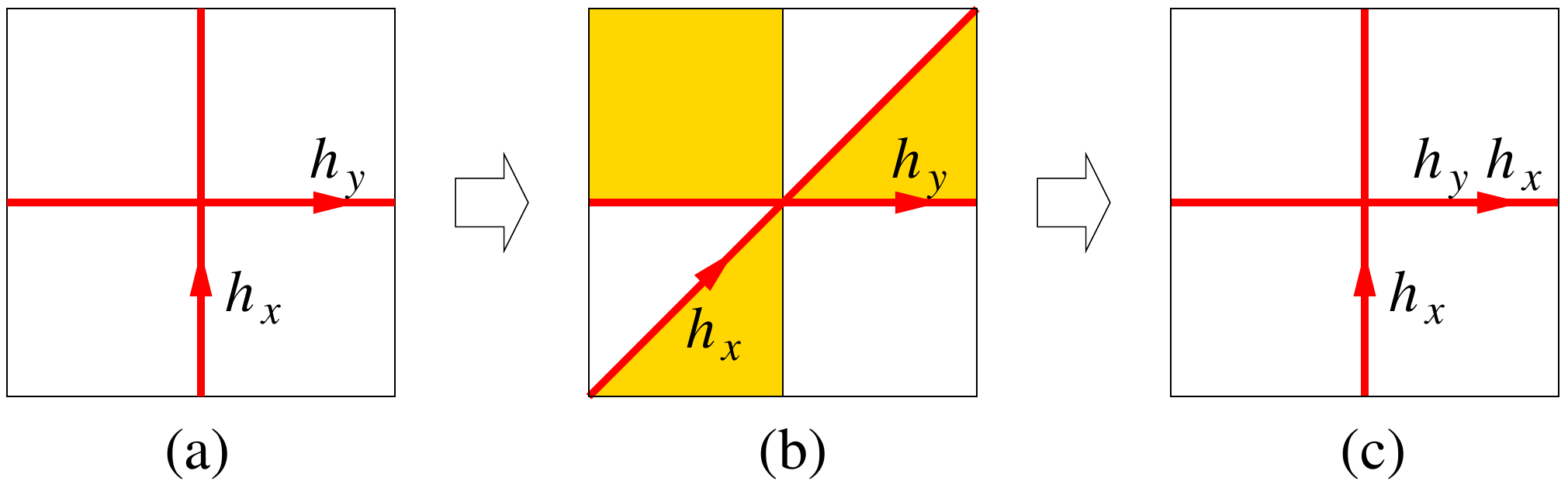} \end{center}
%Fig. 4
\caption{ (Color online) 
(a) A system on a torus with two symmetry twists in $x$- and $y$-directions.
Note that the  torus has the same size $L$ in $x$- and $y$-directions.
(b) The $T$-transformation of the torus, and the
resulting new symmetry twists.
(c) After local symmetry transformation $h_x$ in the shaded region,
the symmetry twists in (b) become  symmetry twists in $x$- and $y$-directions.
} 
\label{Tmov} 
\end{figure}

Using the simulated degenerate ground states and assuming that the Hamiltonian is
translation invariant, we can also simulate the non-Abelian geometric phases in
the topological order.  Let $\Psi_{h_x,h_y}(\{g_{i_x,i_y}\})$ be the wave 
function of $|\Psi_{h_x,h_y}\>$:
\begin{align}
|\Psi_{h_x,h_y}\> = \sum_{\{g_{i_x,i_y}\}}\Psi_{h_x,h_y}(\{g_{i_x,i_y}\})
|\{g_{i_x,i_y}\}\>,
\end{align}
where we have assumed that our system is on a square lattice with periodic
boundary condition, and the physical states on each site, $(i_x,i_y)$, are
labeled by $g_{i_x,i_y}$.  The symmetry twists $h_x,h_y$ are given in Fig.
\ref{Smov}(a) and \ref{Tmov}(a).  

Now let us consider the modular transformations of the lattice
\begin{align}
 \begin{pmatrix}
 i_x\\
 i_y\\
\end{pmatrix}
\to W
 \begin{pmatrix}
 i_x\\
 i_y\\
\end{pmatrix}
,\ \ \ \ W\in SL(2,\Z),
\end{align}
which maps the torus to torus.
The modular transformations is generated by
\begin{align}
 S_0=\begin{pmatrix}
 0&-1\\
 1& 0\\
\end{pmatrix},
\ \ \ \ \
 T_0=\begin{pmatrix}
 1&1\\
 0&1\\
\end{pmatrix}.
\end{align}
The state $|\Psi_{h_x,h_y}\>$ changes under the modular transformation.
Let us define
\begin{align}
 |\Psi^S_{h_x,h_y}\> &= \sum_{\{g_{i_x,i_y}\}}\Psi_{h_x,h_y}(\{g_{-i_y,i_x}\})
|\{g_{i_x,i_y}\}\>,
\nonumber\\
|\t \Psi^T_{h_x,h_y}\> &= \sum_{\{g_{i_x,i_y}\}}\Psi_{h_x,h_y}(\{g_{i_x+i_y,i_y}\})
|\{g_{i_x,i_y}\}\>.
\end{align}

We note that the state $|\Psi^S_{h_x,h_y}\>$ and the state
$|\Psi_{h'_x,h'_y}\>$ have the same symmetry twists if
$(h'_x,h'_y)=(h_y^{-1},h_x)$.  Thus we define a matrix
\begin{align}
\label{tS}
 \hat{ S}_{(h'_x,h'_y),(h_x,h_y)} =
\del_{h'_x,h_y^{-1}}
\del_{h'_y,h_x} 
\<\Psi_{h'_x,h'_y}|\Psi^S_{h_x,h_y}\>.
\end{align}

However,  $|\t\Psi^T_{h_x,h_y}\>$ and $|\Psi_{h'_x,h'_y}\>$ always have
different symmetry twists (see Fig. \ref{Tmov}(b)).  To make their  symmetry
twists comparable, we make an additional local symmetry transformation $h_x$ in
the shaded region Fig. \ref{Tmov}(b), which changes $|\t\Psi^T_{h_x,h_y}\>$ to
$|\Psi^T_{h_x,h_y}\>$. Now $|\Psi^T_{h_x,h_y}\>$ and $|\Psi_{h'_x,h'_y}\>$ have
the same symmetry twists if $(h'_x,h'_y)=i(h_x,h_yh_x)$ (see Fig.
\ref{Tmov}(c)).  Thus we define a matrix
\begin{align}
\label{tT}
 \hat{T}_{(h'_x,h'_y),(h_x,h_y)} =
\del_{h'_x,h_xh_y}
\del_{h'_y,h_y} 
\<\Psi_{h'_x,h'_y}|\Psi^T_{h_x,h_y}\>.
\end{align}

As is evident from the expressions above, for given pair $\{h_x,h_y\}$, it uniquely specifies the values of $\hat{S}$ and $\hat{T}$, since the bra state $\<\Psi_{h'_x,h'_y}|$ with non-trivial overlap with the modular transformed state depends solely on the choice of $|\Psi_{h_x,h_y}\>$. This suggests that we might as well view $\hat{ S}$ and $\hat{ T}$ as some functions of $\{h_x,h_y\}$. i.e. $\hat{ S}_{(h'_x,h'_y),(h_x,h_y)} = F_{\t S}(h_x,h_y)$ and similarly 
$ \hat{ T}_{(h'_x,h'_y),(h_x,h_y)}= F_{\hat{T}}(h_x,h_y)$.
 
Note that both $\hat{ S}_{(h'_x,h'_y),(h_x,h_y)}$ and $\hat{
T}_{(h'_x,h'_y),(h_x,h_y)}$ depend on the size $L$ of the torus (which is the
same in the $x$- and $y$-directions).
Here we conjecture that
$\hat{ S}_{(h'_x,h'_y),(h_x,h_y)}$ and $\hat{
T}_{(h'_x,h'_y),(h_x,h_y)}$ have the forms
\begin{align}
\label{ST}
 \hat{ S}_{(h'_x,h'_y),(h_x,h_y)}&=\e^{-A_S L^2+ o(1/L)} S_{(h'_x,h'_y),(h_x,h_y)}
,
\nonumber\\
 \hat{T}_{(h'_x,h'_y),(h_x,h_y)}&=\e^{-A_T L^2+ o(1/L)} T_{(h'_x,h'_y),(h_x,h_y)}
,
\end{align}
where $A_S$ and $A_T$ are two complex constants (with positive real parts), and
$S_{(h'_x,h'_y),(h_x,h_y)}$ and $T_{(h'_x,h'_y),(h_x,h_y)}$ are topological
invariants that are independent of any local perturbations of the Hamiltonian that
respect the translation symmetry.  This is one of the main results of this
paper. % JH addition of I(g)
 It is possible that 
$S_{(h'_x,h'_y),(h_x,h_y)}$ and
$T_{(h'_x,h'_y),(h_x,h_y)}$, together with the group action operator $I(g)$,
fully characterize the SPT states with a 
symmetry group $G$ in 2+1D.  In the rest of this paper, we will compute $S$ and $T$
for 2+1D  SPT states described by ideal fixed point wave
functions.\cite{CLW1141,CGL1172,CGL1204}

\section{Compute the
universal topological invariants for SPT orders}

In the previous section, we have introduced the notion of symmetry group action matrix
$I(g)$ and modular matrices $S$ and $T$ acting on a basis states containing a pair of twists
on a torus in an SPT phase. In this section, we would like to discuss how these quantities are
computed in idealized fixed point wave-functions describing an SPT phase, 
and the procedure to restore the topologically invariant information stored in these objects. 

\subsection{Twisting the SPT}
It is known that the fixed point wavefunction of a SPT 
is blind to the group cohomology data of the system,
whereas that of the related topological gauge
theory is a very interesting object. The group cohomology data
is revealed in the gauge theory through field configurations consisting 
of \emph{Wilson loops} that
wind non-trivial cycles. This corresponds to link variables 
$\prod_{ij\in \textrm{loop}}\mu_{ij}$, where $i,j,\cdots$ are vertices
along a loop. If this is a flat connection, and that the loop is contractible,
this Wilson loop evaluates to 1. 
These loops are exactly the gauge theory version of the loops introduced in (\ref{loopH}).
The only difference is that in a gauge theory these loops are dynamical degrees of freedom
that can be excited anywhere, whereas in (\ref{loopH}) it appears in designated place.

In the discussion surrounding (\ref{loopH}), the manifold is basically assumed open, such that there is a clear notion of a region inside of the given loop and a region outside. More
generally, say on a torus or more general higher genus surfaces, by insisting that vertices ``inside'' a non-contractible loop gets
transformed by $h$, we are essentially defining a branch cut along the non-contractible loop, and the field configuration is not single valued. 
Such configurations are not
usually considered in the SPT path-integral, because the path integral over field configuration $g(x)$
includes only \emph{single valued} maps $g(x)$ to the target space $G$.
It is however a well-defined link variable from the point of view of the gauge theory.

To reiterate, these twists correspond to ``twisted'' boundary condition
in the field configuration  $g(x)$. 
In particular, in the case of finite group on a lattice, closed loops are represented
by a set of lattice vectors $e=\{e_x, e_y \cdots\}$, which specifies shifts on the lattice
that takes one between identified vertices, a Wilson loop in the SPT phase
can be implemented by
\be
g_{i+ e} =   g_{i}\times h_e.
\ee
This is practically how we specify field configuration on the idealized lattice
wavefunction corresponding to
the transformation effected on the Hamiltonian
described in the diagram (\ref{hxhy}).

The path-integral of the SPT phase now sums over all maps $g(x)$
satisfying the given twisted boundary condition specified above.
Schematically, the SPT path-integral would be given by
\be
\label{SPTpath}
Z_{\textrm{SPT}} = \int \mathcal{D}[g(x)]\vert_{g_{i+ e_\mu} =   
g_{i}\times h_\mu} e^{i S[g(x)]}.
\ee
This however should be contrasted with the path-integral of the
topological gauge theory where all different
gauge configurations or Wilson loops
have to be summed over. In the SPT path-integral the ``Wilson loops''
are twisted boundary conditions, and are thus fixed.
Such boundary conditions would lead to interesting dependence
of the SPT path-integral and ultimately the $S$ and $T$ matrices 
on the group cohomology data, as we will
demonstrate in some simple examples in the following sections.

\subsection{Examples for finite groups}
In order to set the notations for our results, let us begin with
a review of the idealized fixed point wavefunctions on a lattice
describing an SPT with global symmetry group $G$.
\subsubsection{Some preamble on idealized fixed point wavefunctions}
Let us consider the path-integrals, or fixed point wavefunctions 
in greater detail in the case of finite
group $G$ on a discrete lattice. This requires a triangulation
of the space-time manifold $M$ which we denote as $M_{\textrm{tri}}$,
the space-time complex. The action amplitude is then the product of
the amplitude of each $d$- simplex $T_d$ of $M_{\textrm{tri}}$. The
amplitude $\nu_d(\{g_{v_a\in T_d}\})$ on each $T_d$
is a $U(1)$ phase that
depends on the ``field configuration'' $g_{v_a} \in G$
and $v_{a}\in T_d$ denotes the $d$ vertices on $T_d$.
As shown in \Rf{CGL1172} and briefly discussed above,
these phases $\nu_d(\{g_{v_a\in T_d}\})$ satisfy
the $d$-cocycle condition and
the distinct equivalence classes in the
group cohomology group $\mathcal{H}^d(G,U(1))$
corresponds to different phases. To evaluate $\nu_d$
on each simplex for a given representative of an equivalence class
in $\mathcal{H}^d(G,U(1))$,
one needs also to assign a local order of the vertices.
This gives an orientation to each edge and also
determines the orientation of the simplex. Such an order
is called a branching structure. The cocycle is therefore
a function of the field configurations on the ordered
vertices $\nu_d(g_{i_0}, g_{i_1},\cdots, g_{i_d})$, $i_0<i_1\cdots<i_d$.
Such a local ordering can also be obtained if we
assign a global ordering to all the vertices in the complex $M_{\textrm{tri}}$.
As discussed in the previous section, we would like to evaluate the action
amplitude with twisted boundary conditions.

In general, there would be non-trivial cycles in a closed orientable
manifold.
As we described in the previous section, non-trivial closed loops
can be represented by a set of shift vectors $e= \{e_x,e_y,\cdots\}$
on the lattice where vertices related by the shift are identified.
Let us emphasize that these lattice vectors serve to specify
how the multiple (\ie infinite in this case)
copies of the space-time manifold $M$ is identified
and is not related to the lattice discretization in the triangulation
$M_{\textrm{tri}}$.
In this light, the twisted
boundary condition is given by
\be \label{twist}
g_{i + e} = g_{i} h_e,
\ee
for a fixed set of group elements $h_e$ assigned to each of
the $d$ one cycles. The choice for $h_e$ along each 1-cycle
is not completely arbitrary. Very much like the case of Wilson
loops in lattice gauge theory, they have to satisfy consistency
conditions. The basic rule is that the aggregate twist element
along a contractible loop has to be the identity. For example,
on a $d$-torus since $e_\mu + e_\nu - e_\mu - e_\nu$ is a contractible
loop for any $\mu,\nu \in \{1,\cdots,d\}$, it means $[h_{e_\mu},h_{e_\nu}]=1$
for all pairs. This is precisely the same condition already described in section 2. 

To evaluate the action amplitude therefore, we pick a fundamental region
among the multiple copies of space-time manifold introduced.
The boundaries of the
fundamental region are thus $d-1$ dimensional surfaces with normal vectors
given by $e=\{e_x,e_y,\cdots\}$. Action amplitudes of simplices lying well within
the fundamental region can be evaluated as usual. For a simplex that crosses
the boundary with outward normal $\pm e$, the action amplitude is
$\nu(\t g_{i_1}, \cdots \t g_{i_d})$, such that $\t g_{i_n}= g_{i_n}$
for the vertex $i_n$ within the fundamental region, but otherwise
$\t g_{i_n}= g_{i_n} h_\nu^{\pm}$ according to the twisted boundary
condition Eq. (\ref{twist}).

In other words, in a general $d$-manifold $M$,  the twisted boundary condition
above where the field is shifted across a closed loop defined by the shift vector $e$
corresponds to introducing some $d-1$ dimensional branch
surfaces (whose precise positions
are arbitrary and fixed by some chosen convention as in the choice
above for a $d$-torus)
such that the field acquires
a ``twist'' $h_e$ when the cut-surface with normal vector $e$ is crossed.

In order to make direct parallels with the case of lattice gauge theories
we also require that the numbering of the vertex $i_n$ satisfies a twisted boundary
condition
\be
i_n(v_a+ e_i) \to i_n + i\times N_v,
\ee
where $N_v$ is the total number of vertices in $M_{\textrm{tri}}$.
The virtue is that this gives a unique orientation to all
the edges if we stay in one fundamental region of the lattice,
and that edges that are identified in different unit cells would have
consistent orientations.
The path-integral of the SPT phase is now given by
\be\label{SPTpathtwist}
Z_{\textrm{SPT}}(\{h_i\})= |G|^{-N_{v}}\sum_{\{g_{v_a}\}}
\prod_{k} \{\nu_{d}(\{g_{v_a}\}_{T^k_d})\}^{\epsilon_k}
\ee

In the presence of these twists $\{h_i\}$, the path-integral
would depend non-trivially on the cohomology classes.

\subsubsection{$d=3$ }
\begin{figure}[tb] \begin{center} \includegraphics[scale=0.3]{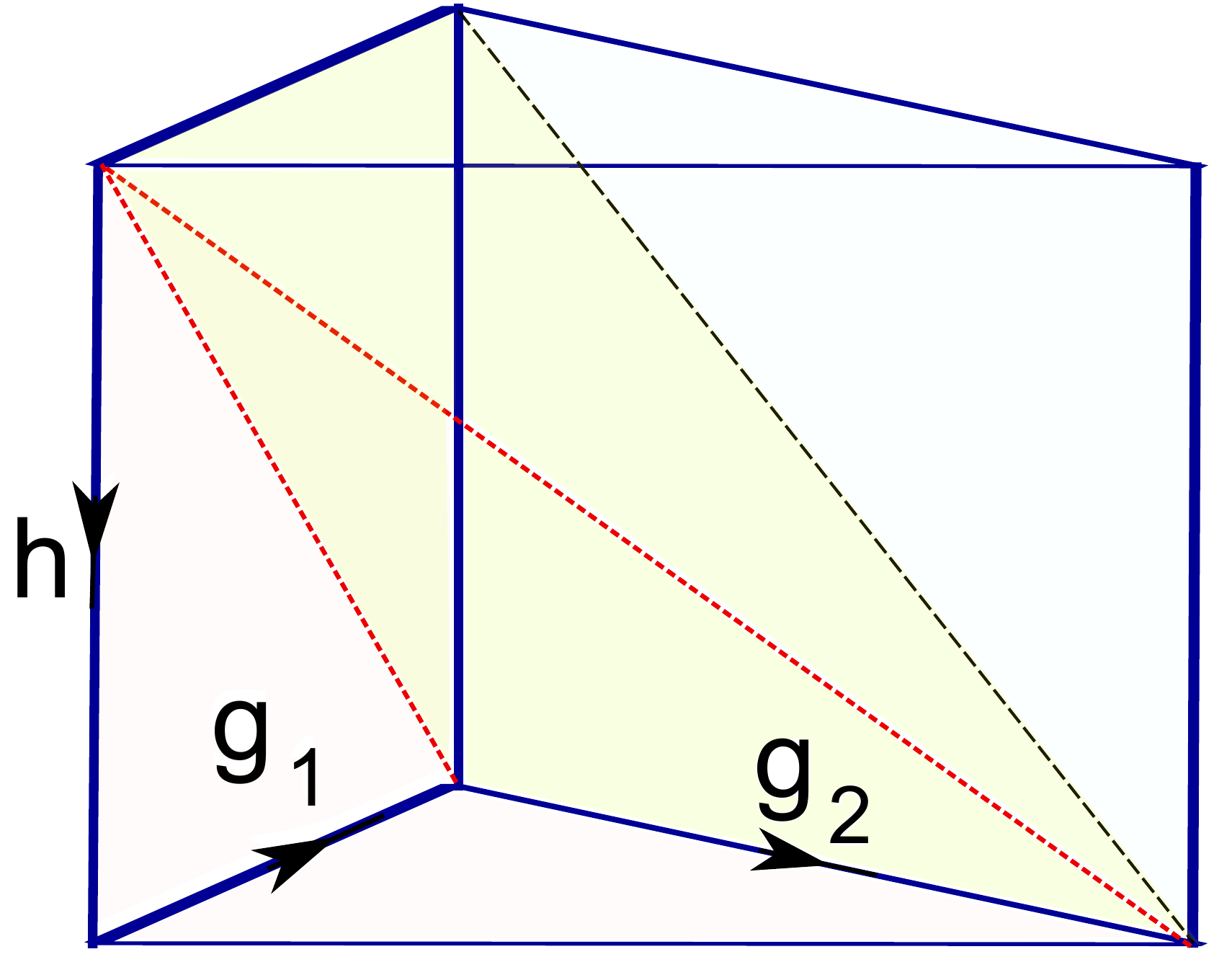}
\end{center}
%Fig. 5
\caption{(Color online) The \emph{cellurarization} into three tetrahedrons of $Y \times S^1$.}
\label{Yblock} \end{figure}

Having discussed very generally the construction of these twisted
path-integrals in general dimensions, let us focus particularly on
the interesting case of $d=3$. 
As a warm up, it is useful to first consider the special case where
the three space-time manifold
concerned is given by $Y \times S^1$, where $Y$ is a 2-sphere with
three holes. The triangulation is represented in Fig. \ref{Yblock},
where the sphere is represented by the triangle whose three vertices are
identified, and the holes by the three edges, and $S^1$ corresponds
to the vertical edge perpendicular to the triangle.

We can again assign twist boundary conditions
on the non-contractible cycles of $Y$.

A group element is assigned to each of the cycles, subjected to the flatness
condition on the triangle. The
consistency condition also immediately follows:
\be
[h,g_i]=1.
\ee
Again, the above condition is a result of
considering the group element assignment to diagonals on any of the
vertical rectangles.
Since the manifold is open, no summation is required over
the group elements, and the path-integral with the orientation
assignment as in the figure is given by 
\be \label{ZY}
Z_{Y\times S_1} = \frac{\alpha(h,g_1,g_2)\alpha(g_1,g_2,h)}{\alpha(g_1,h,g_2)}
\equiv  c_h(g_1,g_2).
\ee
Here for convenience we denote $\nu_3(g_0,g_1,g_2,g_3)\equiv \al(g_0^{-1} g_1, g_1^{-1} g_2, g_2^{-1} g_3) $,
where the cocycles expressed in the form $\al$ is typically used in defining a
lattice gauge theory in which gauge degrees of freedom sit on the links connecting the vertices. 
One important feature to notice is that in the path-integral above the twists on closed loops $h,g_1,g_2$ are not summed over, since they correspond to specific boundary conditions of the field configurations, contrasting what happens in a gauge theory when the link variables are dynamical and has to be summed. 

This combination of three cocycles is denoted $c_h(g_1,g_2)$ because it can be readily shown that it is
a 2-cocycle of the group $N_h$, where $N_h\subset G$ denotes the subgroup
whose elements commute with $h$.
From the relation between $c_h(g_1,g_2)$ and the 3-cocycles
$\alpha$, we should rewrite $ Z_{Y\times S_1}=c_h^{\epsilon}(g_1,g_2) $.
If the orientation of the triangle
aligns with that of the vertical edge we obtain $\epsilon= +1$, and
$\epsilon = \dag$ otherwise.

\subsubsection{Group action and Modular matrices}
\begin{figure}[tb] \begin{center} \includegraphics[scale=0.2]{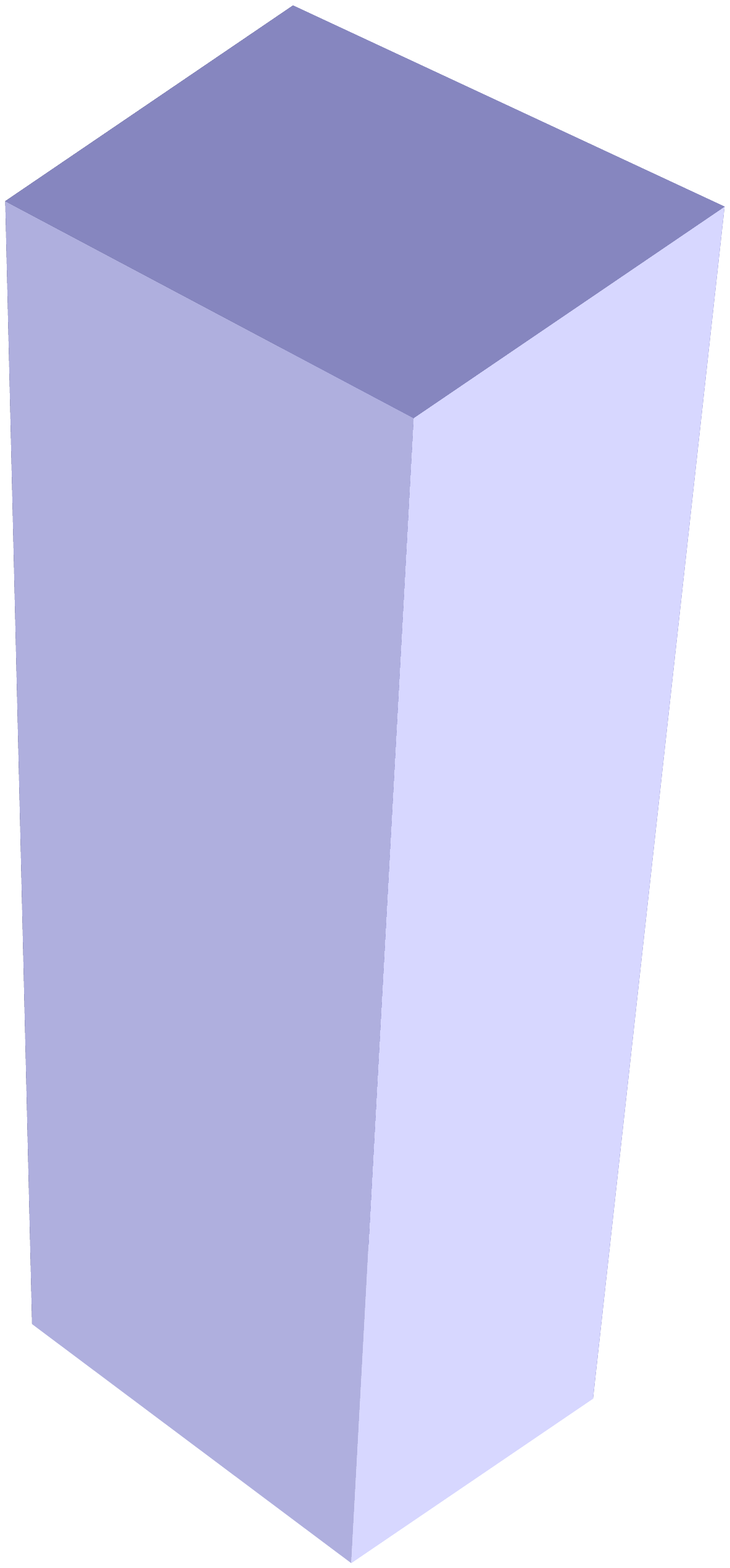}
\end{center}
%
%Fig. 6
\caption{(Color online) An operator acting on a set of basis states defined on a 2d torus. The top and bottom surface rectangles correspond to two states defined on a torus.}
\label{operator} \end{figure}

As is well known that path-integrals evaluated on an open $d$ space-time manifold
with a $d-1$ dimensional boundary surface with specific boundary conditions, we are essentially defining a basis of quantum states in a Hilbert space. Operators acting on this basis of states would appear as a path-integral over a space-time manifold that connects the two different $d-1$ dimensional surfaces where
the quantum state is defined. Geometrically, an operator is therefore a cylindrical object with two different boundary conditions at the top and bottom, and whose path-integral is interpreted as
the matrix element $M_{\Psi_1\Psi_2}  |\Psi_1\rangle \langle \Psi_2|$. 
They are depicted in Fig. \ref{operator}.

\begin{figure}[tb] \begin{center} \includegraphics[scale=0.3]{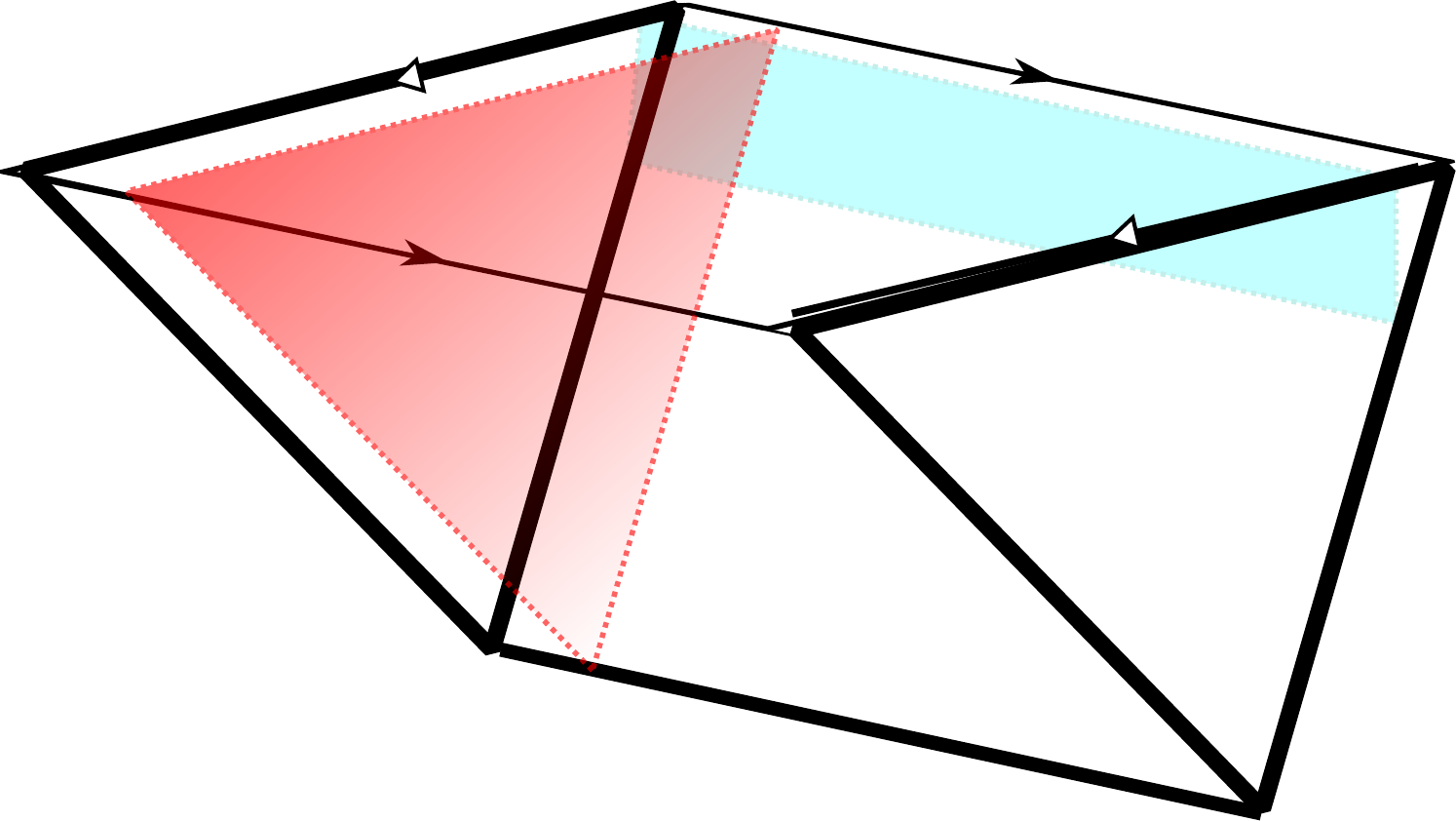}
\end{center}
\caption{(Color online) The depiction of a solid torus. The colored surfaces correspond to
the position of the \emph{branch surfaces} across which a vertex field degree
of freedom is twisted by a group element. The loop on the surface torus that
is contractible in the solid torus extends into a branch surface that ends
in the interior of the solid torus. (blue surface)}
\label{solidtorus} \end{figure}

Now in particular, we can consider placing the system on a solid torus.
The boundary is a two torus which has two non-contractible cycles, although
one of which is contractible in the interior of the torus. See Fig. \ref{solidtorus}.

The path-integral on the solid torus thus defines a basis for the Hilbert
space on the surface torus. In particular, the basis is built up by setting different
boundary conditions on the torus, including different twisted
surface configurations where non-trivial branch cuts are placed on
the boundary. These branch cuts extend into branch surfaces
inside the solid torus and could end in the interior. Since there
are two non-trivial closed cycles, the distinct basis states are specified by the
non-trivial twists along each cycle $|\Psi_{h_x,h_y}\rangle$. In the following,
for simple notations, we denote each state $|\Psi_{h_x,h_y}\rangle$ simply as
\be
|\Psi_{h_x,h_y}\rangle \equiv | h_x,h_y\rangle.
\ee

\begin{figure}[tb] \begin{center} \includegraphics[scale=0.5]{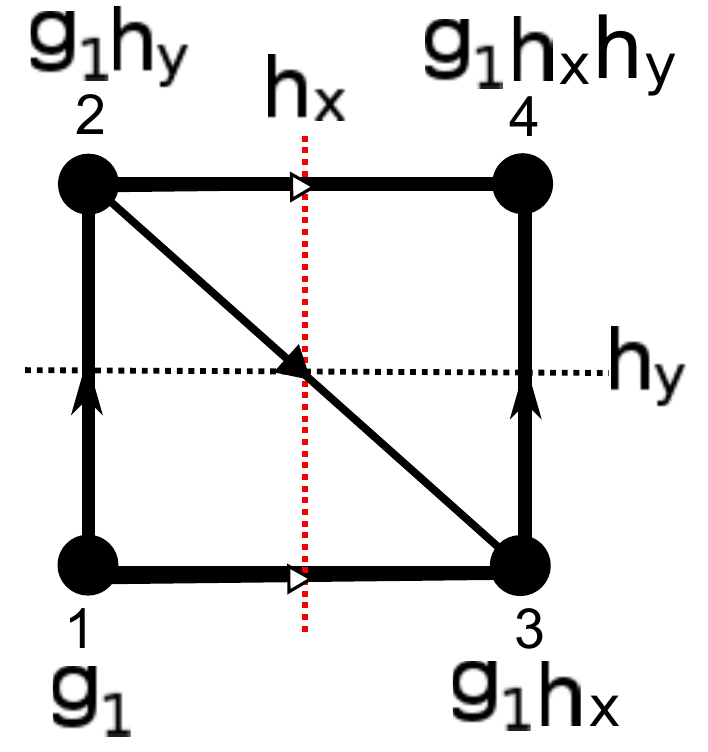}
\end{center}
\caption{(Color online) Our choice of basis states parametrized by different
twists  across branch-cuts (dotted lines) on the surface of a solid torus.}
\label{basistor} \end{figure}

It is a property of the ground state fixed point wavefunction of the SPT phase
defined on an open manifold
that it is insensitive to the triangulation up to a
phase which can be absorbed into the definition of the
wavefunction of the state. This fact will be important as we explore
quantities that are genuinely free of phase ambiguities. Let us in the meantime 
make a convenient choice of our basis states with specific branch cuts on the surface torus by picking
 the simplest possible surface triangulation where there is only one vertex
on the boundary, as depicted in Fig. \ref{basistor}. Vertices labeled 1 to 4
are identified.  Our choice of basis states is
also depicted in Fig. \ref{basistor}. 
\ie
In the picture we have $|\Psi_{h_x ,h_y}(g_1)\rangle$, where $g_1\in G$
is the field degree of freedom sitting at the one vertex of the torus, and $h_x,h_y$
are the twists across the two one-cycle. Note that the path-integral has no dependence on $g_1$.
Note also that in the topological gauge theory, where the symmetry is 
gauged, then each ground state on a torus is given by the sum of a state
characterized by the pair 
 $|h_x,h_y\rangle$ and all its conjugates $|gh_xg^{-1},gh_yg^{-1}\rangle $, where $g\in G$.
This can be thought of as a projection onto $G$-invariant states.
As already emphasized in the previous section, in a SPT the orbits of states under the action of  $G$ 
give rise to a set of physical degenerate states. 

The surface torus is invariant under modular transformation.
Under such a transformation, it is in fact a
reparametrization
of the torus, and thus the Hilbert space has to be invariant
under the transformation, except that our canonical choice of
basis states would be rotated between themselves.

Therefore, we can construct modular transformation matrices
that act on the basis states. These transformation matrices
can be understood geometrically as a cylindrical object that
connect two different solid torus related by a twist. They
can thus be expressed as a product of cocycles.

Let us present here the explicit form of both the $S$ and $T$
generators of the modular group, and also the group action operator $I(g)$.

\begin{figure}[tb] \begin{center} \includegraphics[scale=0.5]{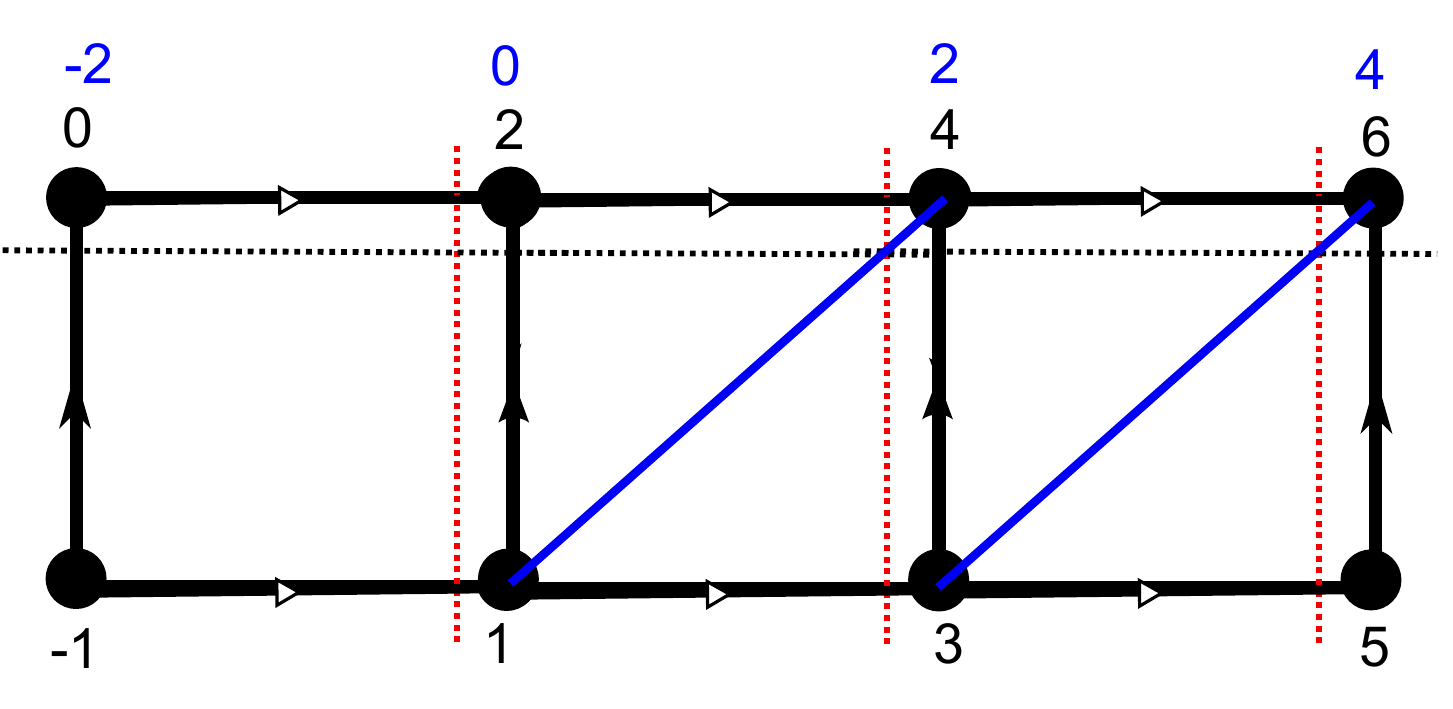}
\end{center}
%Fig. 9
\caption{(Color online) The T-transformation involving a reparametrization of the surface torus.
The blue labels correspond to the new parametrization, and the blue solid lines outline a
parallelogram which is the new choice of unit cell.}
\label{Ttwist} \end{figure}

\begin{figure}[tb] \begin{center} \includegraphics[scale=0.5]{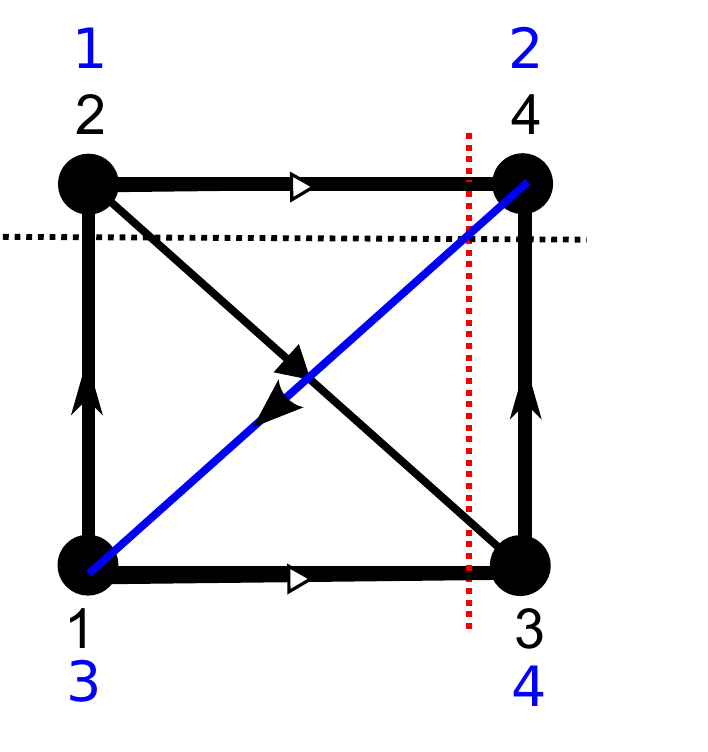}
\end{center}
%Fig. 9
\caption{(Color online) The S-transformation involving a reparametrization of the surface torus.
The blue labels correspond to the new parametrization, equivalent to rotating the torus
by 90 degree. The blue diagonal correspond to the canonical triganulation we have
chosen, now with respect to the new parametrization.}
\label{Stwist} \end{figure}

The $S$-transformation corresponds to rotating the torus, where the complex
structure $\tau \to \frac{-1}{\tau}$.
The $T$-transformation corresponds to a shear transformation where
the complex structure $\tau \to \tau +1$. These transformations
are depicted in Fig \ref{Ttwist} and Fig. \ref{Stwist}.

The $S$-operator, being a path integral on a cylindrical object that connects
the above two parametrizations can thus be triangulated and expressed in terms of
the 3-cocycles.  The same construction appears also in topological lattice gauge theories,
as in \Rf{Yidun}.
We first define an $S$-operator that involves a non-trivial
twist along the \emph{vertical} direction (along the cylinder) leading to branch cuts in the new surface
torus conjugated by element $x \in G$. Choosing $h_x=h, h_y=g$, we have
\begin{eqnarray}
&&S(x) |h,g\rangle = \bigg( \al(hg^{-1},g,x) \al(g^{-1}h,gx,x^{-1}g^{-1}x) \nonumber \\
&&\al(gx,x^{-1}g^{-1}x,x^{-1}hx) \al (x^{-1}hx,x^{-1}g^{-1}h^{-1}x,x^{-1}hx)   \nonumber \\
&&\al (g,x,x^{-1}g^{-1}hx)\bigg)\nonumber \\
&& \bigg(\al(g,g^{-1}h,x) \al(x,x^{-1}hx,x^{-1}g^{-1}x) \bigg)^{-1} \nonumber \\
&&| x^{-1}g^{-1}x, x^{-1}h x\rangle
\end{eqnarray}

Similarly, we can read off the T-matrix with vertical twist $x$ as
\begin{eqnarray}
&&T(x) |h,g\rangle = \bigg( \al(hg^{-1},g,x)\al(hg^{-1},gx,x^{-1}g^{-1}x)\nonumber \\
&& \al(g,x,x^{-1}hg^{-1}x) \al(gx,x^{-1}g^{-1}x,x^{-1}hx) \bigg) \nonumber \\
&& \bigg(\al(g,hg^{-1},x)\al(x,x^{-1}hx,x^{-1}g^{-1}x)  \nonumber\\
&&\al(x^{-1}g^{-1}x,x^{-1}ghx,x^{-1}g^{-1}x)\bigg)^{-1} 
\nonumber \\
&&|x^{-1}hx, x^{-1}ghx\rangle  
\end{eqnarray}

The group action matrix $I(x)$ on the other hand preserves the parametrization
of the surface tori, but introduces a vertical $x$ twist, leading to a conjugation of
the branch cuts on the surface tori. Explicitly, it is given by
\begin{eqnarray} \label{conjugateI}
&&I(x) |h,g\rangle = \bigg(\al(g^{-1}h,g,x)\al(g^{-1}h,x,x^{-1}gx)^{-1}\nonumber\\
&&\al(x,x^{-1}g^{-1}hx,x^{-1}gx)\al(x,x^{-1}gx,x^{-1}g^{-1}hx)^{-1}\nonumber\\
&&\al(g,x,x^{-1}g^{-1}hx)\al(g,g^{-1}h,x)^{-1}\bigg)\nonumber\\
&&|x^{-1}hx, x^{-1}gx\rangle  
\end{eqnarray}

Since the $S(x)$ and $T(x)$ transformation corresponds to a reparametrization of
the torus (in conjunction with shifting the jumps across the branch surfaces by conjugating
by $x$), 
the position of the branch cuts essentially did not
move. They only appear different because we have
made a different choice of the unit cell after the reparametrization. 
Another way to view the transformation is that it maps the three one 
cycles $[v_1v_2]$, $[v_2v_3]$ and $[v_3v_4]$ to another three, $v_1'v_2'$, $v_2'v_3'$
and $v_3'v_4'$ according to the new parametrization as demonstrated in 
Fig. (\ref{Stwist}) and Fig. (\ref{Ttwist}).
We could easily rearrange the branch cuts so that they take the same canonical
shape with respect to the new unit cell,
but the twist is shifted by suitable group elements. 

Note that $S(x)= S(1)I(x)$ and similarly $T(x)= T(1)I(x)$. %as this can
%For Abelian groups, $I(x)=1$, which immediately implies that $S(x)$ and $T(x)$ do not depend on $x$.

\subsection{Caution: Fixing phase ambiguity and true topological invariants}
We would like to pause here and deal with a very important issue: as already
noted above, there are phase ambiguities in our choice of basis states. Therefore
the $I, S$ and $T$ matrix components are generally ill defined quantities!
There are two sources of phase ambiguities. 
First, the three cocycles $\alpha$ are subjected to an ambiguity.
They can be rescaled by a coboundary built on 2-cochains $\beta(g,h)$ as follows 
\be
\alpha'(g,h,k)= \frac{\beta(g,hk) \beta(h,k)}{\beta(g,h)\beta(gh,k)} \alpha(g,h,k).
\ee
Upon rescaling, $S(x)$ and $T(x)$ are rescaled as follows
\begin{eqnarray}
&&\langle g^{-1},h^{-1}|S(x)|h,g\rangle \to 
\langle h,g^{-1}| S(x)|g,h\rangle  \nonumber \\
&& \frac{\beta(g,g^{-1}h)}{\beta(g^{-1}h,g)} \frac{\beta(x^{-1}h^{-1}g^{-1}x, x^{-1}hx)}{\beta(x^{-1}hx,x^{-1}h^{-1}g^{-1}x)}
\end{eqnarray}
and 
\begin{eqnarray}
 &&\langle x^{-1}hx,  x^{-1}ghx|T(x)|h,g\rangle \to 
 \langle x^{-1}hx , x^{-1}ghx|T(x)|g,h\rangle  \nonumber \\
&&\frac{\beta(g,g^{-1}h)}{\beta(g^{-1}h,g)} \frac{\beta(x^{-1}g^{-1}x,x^{-1}ghx)}{\beta(x^{-1}ghx,x^{-1}g^{-1}x)}.
\end{eqnarray}

Similarly,

\begin{eqnarray}
&&\langle x^{-1}gx,x^{-1}hx | I(x) |g,h\rangle \to
\langle x^{-1}gx,x^{-1}hx | I(x)|g,h\rangle   \nonumber \\
&& \frac{\beta(g,g^{-1}h) \beta(x^{-1}g^{-1}hx, x^{-1}gx)}{\beta(g^{-1}h,g)\beta(x^{-1}gx,x^{-1}g^{-1}hx)}.
\end{eqnarray}

This suggests that the surface states is simply rescaled by 
\be
|h,g\rangle \to \frac{\beta(g, hg^{-1})}{\beta(g^{-1}h,g)}|h,g\rangle.
\ee

Second, as noted already earlier, we have picked a particular triangulation
of our tori. For a different choice of triangulation, that corresponds to filling
in extra 3-cocycles $\alpha$. To illustrate that, consider a small change in triangulation
in which we replace the solid line joining vertices 2 and 3 by one which joins vertices
1 and 4 in figure \ref{basistor}.  This change of triangulation would amount to
an extra factor given by $\alpha(h_y, h_y^{-1} h_x, h_y)$, which originates from
fitting an extra three tetrahedral on the surface of the torus. 

The matrix elements  of $I(x), S(x)$ and $T(x)$ therefore generally suffers
from these phase ambiguities, since they correspond to overlaps of
wavefunctions each plagued by phase ambiguities. 

\begin{figure}[tb] \begin{center} \includegraphics[scale=0.6]{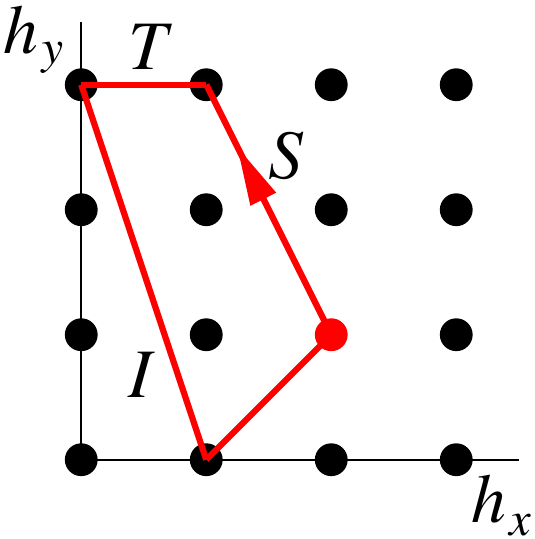}
\end{center}
%Fig. 9
\caption{(Color online) 
$h_x,h_y$ describe the symmetry twists in $x$ and $y$ directions, which label
the simulated degenerate ground states on a torus.  The action of the $S$ , $T$
and $I$ change one simulated degenerate ground state to another.
A closed orbit can be generated by repeated actions of the $S$ , $T$ and $I$
operators.} 
\label{closedorbit} \end{figure}

A natural way to construct invariant quantities is to consider combinations of
$I(x),S(x),T(x)$ that generate a closed orbit in the basis states. As shown in
figure \ref{closedorbit}, a closed orbit connects the same state so that any
phase ambiguity would be canceled out between the state and its own complex
conjugate. One salient example is the combination $(ST)^3$. 
What does this combination of wavefunction overlaps correspond to
from the perspective of path-integrals of topological theories?  This in fact
is precisely the path-integral over a closed manifold. It is well known that a
three-manifold $M$ can be decomposed into two ``handle-bodies'', $M_1, M_2$ of
genus $g$ by cutting $M$ along a genus $g$ surface. This is known as 'Heegard
splitting'.  To reproduce $M$, the surfaces of $M_1, M_2$ have to be identified
in a non-trivial way. In particular, for $g=1$, the non-trivial identification
correspond precisely to doing modular transformations, $S,T$ on the surface
torus of $M_1$ before gluing with $M_2$. Therefore, path-integrals on such an
$M$ in topological quantum field theories take precisely the form
\be
Z_{M} = \sum_{|\Psi\rangle} \langle \Psi | \Gamma | \Psi\rangle, \qquad \Gamma \in SL(2,\mathbb{Z}),
\ee
where $|\Psi\rangle$ is the quantum state defined by the path-integral on the
open manifolds $M_1, M_2$ , and $Z_M$ is the overlap of the same quantum state
after insertion of modular transformation operators $\Gamma$, which are
combinations of $S$ and $T$.  This is indeed what the wavefunctions overlap
computes, in which the $M_1$, and $M_2$ each corresponds to a $T^2 \times
\mathbb{R}$. In the case of topological gauge theories, we project
$|\Psi\rangle$ to a gauge invariant state, by the projector $\sum_{x\in G}
I(x)$, and then finally, we also sum over all the states $|\Psi\rangle$.  In
the case of SPT phase however, no such projection is necessary, and the basis
states with given twists $h,g$ are background configurations that should not be
summed over.  We can thus keep $I(x)$ as an extra operator in our toolbox in
addition to $S$ and $T$, which we could use to operate on our basis states
defined on the surface torus on $M_1$, before taking overlap with the state
defined on the surface of $M_2$.  Therefore, our wavefunction overlaps are
precisely path-integrals of the topological theory underlying the SPT phase
over some closed manifold $M$. 

\subsubsection{$\mathbb{Z}_N$}
Consider specifically $\mathbb{Z}_N$ groups.
The 3-cocycles of $\mathbb{Z}_N$ is given by \cite{MS8977}
\be\label{ZN3cocycles}
\alpha_k(g_1,g_2,g_3) = \exp(\frac{2\pi i k \bar{g}_1}{N^2}(\bar{g}_2
+ \bar{g}_3 - \overline{(g_2+g_3)})),
\ee
for some appropriate $k \in \mathbb{Z}$, and $g_i \in \mathbb{Z}_N$,
and $\bar{x} = x \,\,\,\textrm{mod}\,\, N$ for $x\in\mathbb{Z}$. There are
altogether $N$ distinct choices of $k$ that give rise to representatives
of the $N$ different group cohomology classes in $H^3(\mathbb{Z}_N,U(1))$.
However, this gives $I(x)=1$ identically, independently of the choices of 
branch cuts $h,g$, or the choice of 3-cocycle specified by $k$.

Let us also inspect the form of the $S$ and $T$ matrix of $\Z_N$.
For concreteness, let us look at a few simple cases. 
Substituting into the cocycles, we have, say for $N=2$ 
the following non-vanishing components:

\begin{eqnarray}
&&\left(\begin{tabular}{cc}$\langle 0,0| \,S(1)\, | 0,0\rangle$ & $ \langle 1,0| \,S(1)\, | 0,1\rangle $\\
$\langle 0,1| \,S(1)\, | 1,0\rangle$ &$ \langle 1,1| \,S(1)\, | 1,1\rangle$
\end{tabular}\right) =   \nonumber \\
&&\left(\begin{tabular}{cc}1 & $(-1)^k$ \\
$ (-1)^k$ & $ (-1)^k$
\end{tabular}\right).
\end{eqnarray}

Similarly
\begin{eqnarray}
&&\left(\begin{tabular}{cc}$\langle 0,0| \,T(1)\, | 0,0\rangle$ & $ \langle 0,1| \,T(1)\, | 0,1\rangle $\\
$\langle 1,1| \,T(1)\, | 1,0\rangle$ &$ \langle 1,0| \,T(1)\, | 1,1\rangle$
\end{tabular}\right) =   \nonumber \\
&&\left(\begin{tabular}{cc}1 & $1$ \\
$1$ & $ (-1)^k$
\end{tabular}\right)
\end{eqnarray}

Just to give an example where $g^{-1} \neq g$, we inspect
also the form of  $N=3$, which evaluates to

\begin{eqnarray}
&&\left(\begin{tabular}{ccc}$\langle 0,0| \,S(1)\, | 0,0\rangle$ & $ \langle 2,0| \,S(1)\, | 0,1\rangle $ & $\langle 1,0| \,S(1)\, | 0,2\rangle$ \\
$\langle 0,1| \,S(1)\, | 1,0\rangle$ &$ \langle 2,1| \,S(1)\, | 1,1\rangle$ & $\langle 1,1| \,S(1)\, | 1,2\rangle$\\
$\langle 0,2| \,S(1)\, | 2,0\rangle$&$\langle 2,2| \,S(1)\, | 2,1\rangle$ & $\langle 1,2| \,S(1)\, | 2,2\rangle$
\end{tabular}\right) =   \nonumber \\
&&\left(\begin{tabular}{ccc} 1 & $\exp(\frac{2ik\pi}{3})$ & $\exp(\frac{4ik\pi}{3})$ \\
$\exp(\frac{4ik\pi}{3})$  & $\exp(\frac{2ik\pi}{3})$ &$\exp(\frac{4ik\pi}{3})$ \\
$\exp(\frac{2ik\pi}{3})$ &  $\exp(\frac{4ik\pi}{3})$    &$\exp(\frac{2ik\pi}{3})$ 
\end{tabular}\right).
\end{eqnarray}
Correspondingly,
\begin{eqnarray}
&&\left(\begin{tabular}{ccc}$\langle 0,0| \,T(1)\, | 0,0\rangle$ & $ \langle 0,1| \,T(1)\, | 0,1\rangle $ & $\langle 0,2| \,T(1)\, | 0,2\rangle$ \\
$\langle 1,1| \,T(1)\, | 1,0\rangle$ &$ \langle 1,2| \,T(1)\, | 1,1\rangle$ & $\langle 1,0| \,T(1)\, | 1,2\rangle$\\
$\langle 2,2| \,T(1)\, | 2,0\rangle$&$\langle 2,0| \,T(1)\, | 2,1\rangle$ & $\langle 2,1| \,T(1)\, | 2,2\rangle$
\end{tabular}\right) =   \nonumber \\
&&\left(\begin{tabular}{ccc} 1 & $1$ & $1$ \\
$\exp(\frac{2ik\pi}{3})$  & $1$ &$1$ \\
$\exp(\frac{4ik\pi}{3})$ &  $\exp(\frac{4ik\pi}{3})$    &$1$ 
\end{tabular}\right).
\end{eqnarray}

These quantities, as we discussed, are subjected to rescaling.
There is however a very convenient set of invariants for cyclic groups. Consider
acting the operator $T$ on a state $N$ times, where $N$ is the order of the group--
that necessarily takes us back to the same state.  In the case of $\mathbb{Z}_N$, this gives
\be
\langle h,g| T^N | h,g \rangle = \exp(\frac{2\pi i (h-1)^2 k}{N}), \label{TN}
\ee
where $k$ is the parameter specifying the 3-cocycle as we described above. This quantity turns
out to be sufficient to distinguish all the different $\mathbb{Z}_N$ SPT phases! 
Similar observations can be found in \Rf{Zaletel}.
The combination $(ST)^3$ however evaluates to 1 always.

\subsubsection{$\mathbb{Z}_N\times\mathbb{Z}_N\times \mathbb{Z}_N$}

The group elements of the group is denoted by a ``three-vector'' $g=
(g_1,g_2,g_3)$.  The cohomology group $\mathcal{H}^3(\mathbb{Z}_N^3,U(1))$ has
seven generators.  Six of them involves only two of the three $\mathbb{Z}_N$,
which lead to trivial results. The interesting generator intertwines the three
$\mathbb{Z}_N$. The corresponding topological gauge theory contains non-Abelian
anyons.  This set of cocycles takes the following form:
\be\label{Zn3gen}
\alpha_k(h,g,l)= \exp(\frac{2\pi k i}{N} h_1g_2l_3).
\ee
Note that the action of $I(x)$ becomes $x$ dependent for this set of cocycles. 

Evaluating on $T^2\times S^1$ gives
\begin{eqnarray}
&&Z_{T^2\times S^1}(h,g,x)_{\al_k} = \exp(\frac{2\pi k i}{N}(h.g\times x))  \nonumber\\
&&=  \langle h,g| \,I(x)\,|h,g\rangle,
\end{eqnarray}
where $[x,g]= [x,h]=[h,g]=1$, and $h.g\times x= \epsilon_{abc} h_a g_b x_c$.
In this case therefore, we find that each element of $I(x)$ is a topological
invariant capable of distinguishing all these non-trivial 3-cocycles. 

We can also evaluate $S(1)$ and $T(1)$. In fact, for $N=2$, $T(1)$ simplifies
to
\be
\langle h, gh |T(1)| h,g \rangle = \exp\left(\pi k i (h_1g_2g_3+h_2g_1g_3+ h_3g_1g_2)\right).
\ee
Similarly, $S(1) $ is given by
\begin{eqnarray}
&&\langle g^{-1}, h |S(1)| h,g \rangle=\exp\bigg(\pi k i (g_1g_2g_3  +\nonumber\\
 &&h_1h_2h_3 + h_1g_2g_3 +
g_1g_2h_3 + h_1g_2h_3)\bigg).
\end{eqnarray}

In this case of course the matrix elements of $T^N$ are again topological
invariants. With $I(x)$ and the matrix elements of $T^N$ one can distinguish
all possible SPT phases with  $\mathbb{Z}_N \times
\mathbb{Z}_N\times\mathbb{Z}_N$ symmetries.

\subsubsection{Some examples of non-Abelian groups}

Let us also inspect some simple cases with non-Abelian symmetry groups.  In
particular, the simplest such example is the Dihedral group $D_N$, for $N$ odd.
The $2N$ group elements can be represented as a pair $(A,a)$, where
$A\in\{0,1\}$, and $a \in \{0,\cdots,N-1\}$. Group product is given by 
\be
(A,a) \times (B,b) = ( (A+B)_{\textrm{Mod}\, 2}, ((-1)^B a+b)_{\textrm{Mod}\, N}.
\ee
The group cocycles are given by
\begin{eqnarray} \label{DNcocycle}
&&\al_k (g,h,l) = \nonumber \\
&&\exp\bigg(\frac{2\pi i k }{N^2}\big((-1)^{H+L} g[ (-1)^L h+l -((-1)^L h +l)_{\textrm{Mod}\,N}]    \nonumber\\
&&+ \frac{N^2GHL}{2}\big)\bigg),
\end{eqnarray}
where as explained above a group element $g$ corresponds to the pair $(G,g)$,
and with an abuse of notation we use the same symbol $g$ also for the second
component in the pair, where it is $\in \Z_N$. 

Evaluated on $T^2\times S^1$, one again requires that the monodromies across
each of the branch-cuts, given by $h,g,x$ have to be mutually commuting. In the
Dihedral groups where $N$ is odd, two elements $g_1$ given by $ (G_1,g_1)$, and
$g_2$, corresponding to  $ (G_2,g_2)$ are mutually commuting if and  only if
they belong to one of the following situations: (1) they are both in the $Z_N$
subgroup where $G_1=G_2=0$; (2) if $g_1$ and $g_2$ are the same group element;
(3) either $g_1$ or $g_2$ is the identity element $(0,0)$. As a result
\be
\frac{\al_k(g,h,l)}{\al_k(g,l,h)} = 1 
\ee 
for any $k$ if $g,h,l$  are mutually commuting. As a result, it is also clear
from Eqn. (\ref{conjugateI}) that the partition function on $T^2\times S^1$ is
unity for any set of allowed branch cuts.

We could also look at some examples of $I(x)$.  It is not hard to check that
for odd values of $N$
\begin{eqnarray}
&&\langle x^{-1}hx, x^{-1}gx|I(x)|h,g\rangle =
\exp\bigg( \frac{2\pi k i (-1)^{G+X}}{N^2}\big[\nonumber \\
&&(g^{-1}h) <(-1)^X g + x(1-(-1)^G)> - \nonumber\\
&&(-1)^H g <(-1)^X g^{-1}h + x(1-(-1)^{G+H})>\big]
\bigg),
\end{eqnarray}
where the pointed bracket above means $< n> = n - (n)_{\textrm{Mod}\,N}$.  The
$T(1)$ matrix can be simplified to
\be
\langle h,gh|T(1)|h,g\rangle = \al(g,g^{-1},h)\al(g^{-1},h,g)\al(h,g,g^{-1})
\ee
and the corresponding $S(1)$ can be massaged to the form
\be
\langle g^{-1},h| S(1) | h,g\rangle = \al(h,g^{-1},h)\al(g^{-1}h,g,g^{-1}h)\al(h,g^{-1}h^{-1},h).
\ee

In this case, the operator $S(1)$ acting on basis states with $h=g=\{1, a\}$
already forms  closed orbits. Therefore each matrix element, which evaluates to
the following 
\be
\langle h,g| \hat{S}(1)|h,g\rangle\vert_{h=g=(1,a)} = (-1)^k,
\ee
is a topological invariant. 

This only distinguishes $k$ even from $k$-odd 3-cocycles.
Similar to cyclic groups, one can also inspect
$T^N$, acting on states $|h,g\rangle$ such that
$h,g$ lives in the $\mathbb{Z}_N$ subgroup, taking the form
$  \{0,a\}$. In this case,
$T^N$ evaluates precisely to the same value as in
(\ref{TN}), with $h$ replaced by $a$ for the element $h=\{0,a\}$.
Here $(ST)^3$ again evaluates to 1 identically.

\section{Summary}
In this paper, we propose a systematic way to construct `order parameters' of
SPT phases, by exploiting the relationship between an SPT phase and the
corresponding intrinsic topological order obtained by gauging the global
symmetry described in \Rf{LG1220} .

To simulate the effect of gauging, the idea of the symmetry twist is
introduced. Symmetry twists are generated by symmetry transformations performed
in a restricted region. The boundary of such a region would play the analogous
role of a Wilson line in a topological gauge theory, although in contrast to
Wilson lines in a gauge theory, these twists are not dynamical excitations, but
background defects. 

In the special case of a 2+1 d SPT we can consider putting the system on a
torus, \ie one that satisfies periodic boundary conditions.  States with
symmetry twists wrapping the two cycles on the torus can be constructed,
leading to a set of almost degenerate basis states. This is again analogous to
the situation of a gauge theory with non-trivial Wilson lines.  Modular
transformation on a torus, corresponding to a reparametrization of the lattice
on the torus, can be performed, which would take us to a different basis state,
so does a global symmetry transformation.  One could consider overlaps of the
wavefunctions between a state $|\al\>$ with some given symmetry twists and a
modular and global symmetry transformed version of another state $|\bt\>$:
\begin{align}
\<\al| \hat U |\bt\>
= \text{e}^{-L^2/\xi^2+o(1/L)}
 U_{\al\bt}. 
\end{align}
We conjecture that the factor $U_{\al\bt}$ is universal which can be used to
characterize the SPT states.

These wavefunction overlaps can be understood as path-integrals of the SPT
phase on a three manifold with two boundaries, where each boundary is our
torus, which is a constant time slice on which we define our states.  These
wavefunction overlaps therefore generally suffer from a phase ambiguity, since
they are not invariant upon rescaling each basis state with an arbitrary phase.
To construct truly universal topological invariants, we further consider
closed orbits of the modular transformations. \ie We systematically look for
combinations of these modular transformations and global symmetry
transformations on our basis states such that it keeps the states invariant up
to an overall phase. 

These  overall phases generated by closed orbits of transformations are the
topological invariants that we set out to find: like Berry phases, they are
independent of any rescaling of our basis states by arbitrary phases. We
demonstrate how in various simple symmetry groups, both Abelian and
non-Abelian, such closed orbits can be constructed and how they can be
extracted using idealized fixed point wavefunctions. It is amusing that, as
explained in the text, these special wavefunction overlaps are equivalent to
path-integral in some  highly non-trivial  \emph{closed} manifold. In the
examples we explored in detail, the precise combination of modular
transformations leading to closed orbits are generically dependent on the
symmetry group involved, and in all the discrete symmetry groups we have
considered, there are enough closed orbits that distinguish all the distinct
SPT phases with the given symmetry classified in \Rf{CGL1204}.  This supports
the conjecture that our constructed topological invariants fully characterize
the SPT states.

LYH would like to acknowledge useful discussions with Yidun Wan. 
This research is supported by NSF Grant No.  DMR-1005541, NSFC 11074140, and
NSFC 11274192.  It is also supported by the John Templeton Foundation.
Research at Perimeter Institute is supported by the Government of Canada
through Industry Canada and by the Province of Ontario through the Ministry of
Research. The research of LYH is supported by the Croucher Foundation.

\bibliographystyle{apsrev}
%\bibliography{TensorNets}% Produces the bibliography via BibTeX.
\bibliography{../../bib/wencross,../../bib/all,../../bib/publst,./local}

\end{document}